# PASSWORD BASED A GENERALIZE ROBUST SECURITY SYSTEM DESIGN USING NEURAL NETWORK

**Manoj Kumar Singh**
**Manuro Tech. Research Lab, Bangalore, India**
**mksingh@manuroresearch.com**

**Abstract**
Among the various means of available resource protection including biometrics, password based system is most simple, user friendly, cost effective and commonly used. But this method having high sensitivity with attacks. Most of the advanced methods for authentication based on password encrypt the contents of password before storing or transmitting in physical domain. But all conventional cryptographic based encryption methods are having its own limitations, generally either in terms of complexity or in terms of efficiency. Multi-application usability of password today forcing users to have a proper memory aids. Which itself degrades the level of security. In this paper a method to exploit the artificial neural network to develop the more secure means of authentication, which is more efficient in providing the authentication, at the same time simple in design, has given. Apart from protection, a step toward perfect security has taken by adding the feature of intruder detection along with the protection system. This is possible by analysis of several logical parameters associated with the user activities. A new method of designing the security system centrally based on neural network with intrusion detection capability to handles the challenges available with present solutions, for any kind of resource has presented.
*Keywords: artificial neural network, security, authentication, intrusion detection*.

## 1. Introduction

Security is a broad topic and covers many issues. Malicious people trying to gain some benefit, attention, or to harm someone intentionally cause most security problems. As, Complete security not possible in real life, transition will be long in coming. Conventional cryptographic methods have their own problems. Intrusion Detection System (IDS) is an emerging new technology, being informed is the best weapon in the security analyst's arsenal. "An ounce of prevention is worth a pound of detection". An Intrusion Detection System detects attacks as soon as possible and takes appropriate action. Security is a compulsory need for data operation today. Information or commerce exchanges need security and reliability. Examples are like: Bank Transactions where state–of-the art financial security is mandatory, Protection of Personal Resources etc. password based security system should must posses facilities like: (i) provision to assign complex password without any restriction. (ii) Password must be encrypted up to proper level before storing or transmitting. (iii) Easy and secure means to reset the password. (iv) Facility to record and monitor failed login attempt. (v) Don't let people try to break password in forever. General structure of security system is shown in fig. (1), to access the resource user must verify by authentication environment.

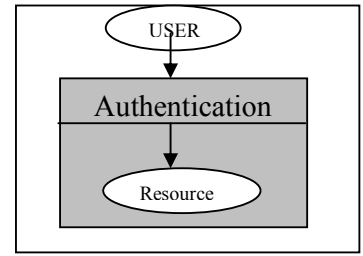

**Figure 1. Security System in Abstract form**

Process of authentication can be defined as developing a unique mapping process from given password to some other unique information in a defined domain. The guarantee of security doesn't only depend on unique mapping but greatly depend upon difficulties associated with getting back the password form the mapped formation. Artificial Neural network characteristics, making it very appropriate for playing an important role in coming era of security design. Artificial Neural Network **(ANN)** is an information-processing paradigm that is inspired by the way biological nervous systems, such as the brain, process information. The key element of this paradigm is the novel structure of the information processing system. It is composed of a large number of highly interconnected processing elements (neurons) working in unison to solve specific problems. ANNs, like people, learn by examples. An ANN is configured for a specific application, such as pattern recognition or data classification, through a learning process. Learning in biological systems involves adjustments to the synaptic connections that exist between the neurons. This is true of ANNs as well.

## 2. Literature Review

In order for a computer system to perform specific acts on behalf of a specific individual, an identification and authentication step is needed. This process of identification and authentication has been extensively studied and formally described in principle [1], [2]. These formal descriptions include the user as the principal and document all relevant constructs and issues from a single system perspective. In the basic authentication process, the entity desiring authentication presents credentials usually an account ID and some additional information, to prove that the request is coming from a legitimate owner of the ID.This is relatively straightforward process that has been in use for decades. An example is user ID and password combination, one of the simplest forms of user authentication [3]-[6]. A more complicated example is the smart card system [7], [8], where a user typically has an ID, a password and also a dynamic (time-generating) passkey from the smart card which changes every 60 seconds. The authenticating server has the same time changing numerical sequence as the specific smart-cards assigned to that ID and if the ID, password and card generated number are all correct, authentication is granted. Frequently smart-card are combined with passwords for an account to increase security. This is an example of two-factor authentication and is more secure because it requires more items for authentication. Another form of authentication involves the concept of biometrics. Biometrics can take the form of several measurements, from fingerprints to retinal scans to pupil images. The advantage of biometrics is that always available with user without any extra means and effort and unforgotten. Disadvantages are many, including not being able to change them if needed and complexity of solution. Many modern system have adopted a simple ID/password method of achieving the goals associated with the identification and authentication function, and numerous technical method exists to achieve this [3],[4],[7],[9],[15].the wide Varity of implementation schemes are a result of individual design decisions appropriate to specific circumstances at the time to design of a specific system or class of a system. Important research on human cognitive ability has generated a lot of practical knowledge on the issue of what an individual can remember [16]. The effect of human cognitive ability in the authentication process is a central element, through often overlooked by developers. Remembrance of passwords is one of the cornerstones of the current password based authentication system [17],[18],[24].From system perspective a password should be easily remembered, yet hard for an intruder to guess[19]-[21].Although other system level solutions exist much of the effort to secure password-based system is focus on preventing unauthorized access through better password selection[5],[13],[29],[30].Many known weaknesses exists in password-based system and various fixes have been applied over time[4],[9],[10],[11],[12],[14].

According to the learning ability, artificial neural network has been used in artificial intelligence [31], neural network can be used to model nonlinear statistical data, which can model complex relationship between inputs and outputs. Sometimes, it can generate chaos phenomenon. According to this property, it has complex dynamic action, which can be used to protect data content. For example, the random sequence produced by neural network can be used to encrypt data [38], and the neural networks that generate chaos phenomenon can be used in secret communication [39]. According to complex relation between the nodes of neural network, neural network can produce the sequences with random properties. For the neural network that has chaotic dynamics, its output is often sensitive to the inputs or such control parameter as weight; it is caused by parameter sensitivity of chaos system. Taking the neural network proposed in [37] for example, although there is a slight difference in the initial value, the output changes greatly. This property makes the initial value suitable for the key that controls the data encryption or decryption. Application of neural network for intrusion detection has been shown in [36]. Intrusion detection is an important technology in network security, which can detect illegal intruders or illegal intrusions. Using neural network's supervised learning, the intrusive operation can be distinguished from normal operation [37]. The one-way property makes neural network a suitable choice for hash function. Hash function is a technique for data integrity authentication. Till now some hash functions based on neural network have been presented [38][39], which were reported to have some advantages compared with existing schemes, such as high time efficiency or flexible extension.

## 3. Risk and complications associated with password authentication

### 3.1 Attacks on password

Because of high sensitivity with attacks, risk factor is very high in password based authentication process. Broadly the types of attack can be divided into three categories: - technical (brute force), discovery and social engineering.

In the brute force attack, two methods can be used, (a) attempting passwords against the system, but this is easily stopped with account lockouts. (b) An offline attack against the password hash file. This is processor intensive search through the entire password key-space, calculating and comparing hash values of potential passwords to the values in the stolen hash file. Various defense exist, including increasing key-space through the use of salts, and physically protecting the password hash file.

Password may also be compromised by discovery. Forms of password discovery may vary and include interception of a script file, an exploit on another system, a Trojan program capturing keystrokes, or the discovery of default passwords associated with other system or programs. The primary defense against discovery is proper system design rules that do not allow discovery of passwords through scripts or default system accounts.

Social engineering represents an attempt by an intruder to elicit password and account information from a user. This attack is exogenous to the computer system in question, coming via phone, fax, email or casual contact. The primary defense against this type of exploit is training and awareness directed at the user with respect to this specific vulnerability.

### 3.2 Complications with safeguards

Password rules are either optional or enforced specification about the length of the password and the diversity of the characters that comprise its' the length and diversity contribute to the size of the domain set containing all passwords that increase the difficulty of brute force detection. Prevention of easily guessed passwords reduce discovery. However the same rules that increase password resistance to brute force attack directly reduce discovery. However the same rules that increase password resistance to brute force attack directly reduce the ability of a user to remember a password and increase the need of password aids. System rules relate to the procedural aspects of gaining access and are enabled in a system. For example, the automatic user lockout after three failed trailed attempts is a system-enforced rule. More complex mechanism includes expiring passwords and the forcing of password changes. System rule can also have the opposite effect through, as they can lead to discovery patterns. There are systems that will email an encrypted password back to user if required presenting an opportunity for discovery. System rule that make the passwords harder to remember can increase the need for user based password memory aids system rules for recovering forgotten password such as emailing forgotten passwords lessen the need for memory aids but increase risk of discovery. In general rules that enforce higher quality passwords do so at the expense of user memory ability. This was an acceptable trade off when the count of passwords per user was low but this trade off today forces users to use memory aids. One of the weakest links in a security system is an untrained user. Formal and informal activities of training and awareness can alleviate a wide Variety of actions that weaken a system such as choosing poor passwords, writing them down, sharing with others. Training can address issues associated with discovery and social engineering attackers. The primary issue with training is its temporary nature, users forget or become complacent over time and retraining is time consuming and costly.

### 3.3 Complications because of multi-usability

Today users have multiple accounts on multiple systems. users must to remember multiple IDs and multiple passwords for wide range of computer based service they use, this has placed a strain on user memory and users have developed memory aids, such as password lists to assist them in the task of keeping accounts and passwords straight. User password memory aids affect overall system security at the individual application level.

## 4 Neural network's features suitable for security design

### 4.1 Learning ability

Neural network has the possibility of learning. Given a specific task to solve and a class of function, neural network can use a set of observations to solve the task in an optimal sense. Generally, according to learning task, learning ability of neural network can be classified into two main categories, i.e. supervised learning, and unsupervised learning. Supervise learning is the learning with a 'teacher' in the form of a function that provides continuous feedback on the quality of solutions. These tasks include pattern classification, function approximation and speech recognition, etc. Unsupervised learning refers to the learning with old knowledge as the prediction reference. These tasks include estimation problem, clustering, compression or filtering.

### 4.2 One-way property

One-way property means that it is easy to compute the output from the input while difficult or impossible to compute the input from the output. In neural network, there are often more inputs than outputs. Taking a simple neuron model for example, the input is composed of n elements $P_0, P_1, P_2...P_{(n-1)}$, while the output is a unique element C. it is defined as

$$C = f \left( \sum_{i=0}^{n-1} w_j p_j + b \right).$$

As can be seen, it is easy to compute C from $P=[p_o, p_1, p_2...p_{(n-1)}]$, while difficult to compute P from C. The difficulty is equal to solve a singular equation. Thus it is one-way process from the input P to the output C. This property is often required by Hash function [35][36] that is used to authenticate data's integrity. This is just one example with one neuron model, complexity of getting C from P with network of neurons can be understand.

## 5 Proposed authentication method based on neural network

As discuss earlier, the two main requirements for developing authentication system are, unique mapping and no reversibility of mapping. These two objectives can be fulfilled by means of applying the suitable learning of password and developing an appropriate architecture. Feed forward architecture has taken, over that the user-defined password can be trained with supervise learning algorithm. This section will contain details about architecture of neural network, learning rule, target definition and process, which will apply for authentication.

### 5.1 Feed forward architecture

The type of Architecture implemented in the project is multiplayer Feed Forward architecture as shown in Fig (2). This neural network is formed in three layers, called the input layer, hidden layer, and output layer. Each layer consists of one or more nodes, represented in this diagram by the small circles. The lines between the nodes indicate the flow of information from one node to the next. In this particular type of neural network, the information flows only from the input to the output.

The nodes of the input layer are passive, meaning they do not modify the data. They receive a single value on their input, and duplicate the value to their multiple outputs. In comparison, the nodes of the hidden and output layer are active. This means they modify the data as shown in Fig (3) each value from the input layer is duplicated and sent to all of the hidden nodes.
The values entering a hidden node are multiplied by weights; a set of predetermined number is stored in the program. The weighted inputs are then added to produce a single number. This is shown in the diagram by the symbol, **Σ**. Before leaving the node, this number is passed through a nonlinear mathematical function called a Sigmoid. This is an **"S"** shaped curve that limits the node's output. That is, the input to the sigmoid is a value between $-\infty$ to $+\infty$, while its output can only be between 0 and 1. The outputs from the hidden layer given to output nodes after multiplication with associated weight on that path. The active nodes of the output layer combine and modify the data to produce the output value of the network. In the case of this project the output layer only needs a single node, single hidden layer contains number of nodes equal to 30 % of the size of the input layer nodes number. The input layer contains the number of nodes according to length of the user identity. The initial weights are generated randomly using uniform distribution within the range of [0 1].

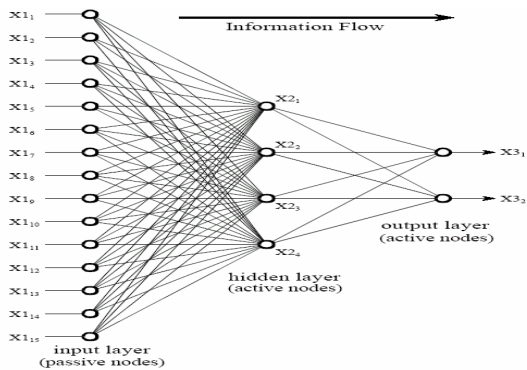

**Figure 2. Feed forward architecture**

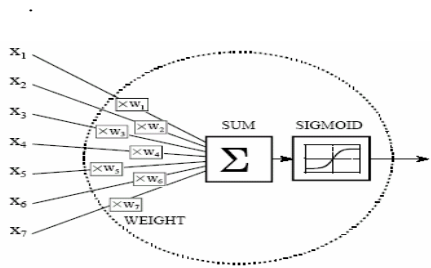

**Figure 3. Neural Network Active Node**

### 5.2 Transfer function: Sigmoid Function

Sigmoid function (unipolar), mathematically described by the equation $S(x) = 1/(1+e^{-\lambda x})$, having several advantages like, (1) Soft limiter, i.e. having sensitivity w.r.t variation in input. (2) Mathematical model of biological neuron, firing phenomenon the characteristic is appearing like sigma function. (3) It derivative is easily available which is required in learning process, $S'(x)= s(x)[1 - s(x)]$.

### 5.3 Physical interpretation of learning rule

Mathematical modeling of any physical phenomena has at least two purposes like Compact form representation and simplification in processing. But sometimes when we observe physical phenomena properly, there may be a chance to get the same solution as it comes from the mathematical modeling.

In ANN, the main goal of learning rule is to minimize the *error*, which is the difference between the target and the observed network output. Weight training is usually formulated as minimization of an error function, such as the mean square error between the target and the actual outputs averaged over all by iteratively adjusting connection weights.

**Why this error exists?**

Error exists because initially there is zero co-relation between the connection weights and the applied inputs, since the connection weight values are chosen to be random. So the error value is high initially.

**How this error can be minimized?**

In simplified form we can say, if error is high, it indicates that the current weights are very far from the required value and hence a large change in weights required. In other case if error is less, it indicates that the current weights are near to the required value and hence change in weights required for the next iteration is also small. Hence we can say, $\Delta W \propto$ Error. Each and every weight is responsible for the output, but in different proportion. So it seems logical to adjust the individual weights in accordance to their affect at the output in order to minimize the error, where the affect is calculated by taking slope, i.e., variation at the output w.r.t the variation in individual weight. Hence we can say, $\Delta W \propto$ Affect**.** Combining with placement of proportional constant η, the final equation of weights adjustment can be expressed as $\Delta W=\eta(Error)(Affect)$. Advantage of this method for finding the change in weights are simplicity and very suitable for more complex architecture (like neural network with two or more hidden layers. Author claims to have a method for finding weights adjustment equation for any complex circuit just by visual appearance of network).

### 5.4 Target selection for high sensitivity

Because of supervise learning method; target must be defined at beginning. The output node transfer function is a sigmoid function; hence any value on the curve of sigmoid function can take as target i.e. between 0-1.Authentication system should must be very sensitive w.r.t even a very small variation in the input. The sigmoid characteristic curve can be divided in the three regions namely regions (1)&(3) can be consider as saturation region, because in these regions even for some good variation in input of sigmoid function, o/p variation is not very much. Region (2) can be considered as linear region, because changing in o/p with input is approximately linear manner as shown in fig (4). To enhance the sensitivity it's necessary to define the target value in the linear region and the most suited value is equal to 0.5.

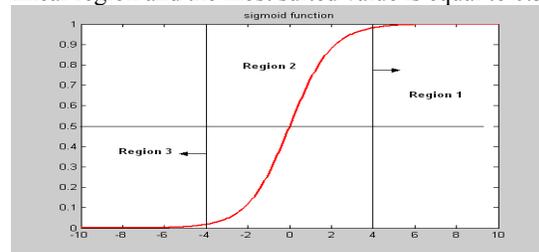

**Figure 4. Sigmoid curve with 3 regions**

## 6. Authentication process

Unique mapping of a given password can achieve by giving the training of neural network with taking password as an input. Because row information of password may have various characteristics like having alphabet character, numeric or even some special character or any combination. Hence it's not possible to go for direct training. A preprocessing step required transforming the row data in some suitable format like numeric only. Sensitivity with authentication can be further improved very much if row password is transformed in to corresponding binary format. According to length of this transformed password an architecture with same number of input nodes, one hidden layer containing around 30% nodes of input layer and one output node created. To get the maximum sensitivity this architecture is fully interconnected. Sigmoid function has taken as transfer function of any active node. Once training completed up to specified error value, the final output of network is the final mapped valued of the password. The hidden nodes output also consider as the intermediate mapped values. Even final mapped value is alone enough to provide the authentication but inclusion of intermediate mapped values will give added strength of complete authentication process**.**

　　Any test case, first pass through the same preprocessing step as original password has been. If length size of test password is same as original password (otherwise un-authentication declared), same architecture taken as in case of original password, along with trained weights (learning will not take place this time). With the processing of architecture, all mapped values (final & intermediate) compared with the all-original mapped values. Zero difference in result counted as authenticated action otherwise counted as an unauthenticated.

| Correct password: architecture | |
|---|---|
| Architecture size | |
| Layer | No. Of Nodes |
| Input | 84 |
| Hidden | 25 |
| Output | 1 |

## 7.Experimental results

According to specified description of neural network and its utilization of authentication, several experiments have done. Two examples along with some unauthenticated trail have shown below.

### 7.1 Experiment 1

| Correct password: neural | |
|---|---|
| Architecture size | |
| Layer | No. Of Nodes |
| Input | 42 |
| Hidden | 13 |
| Output | 1 |

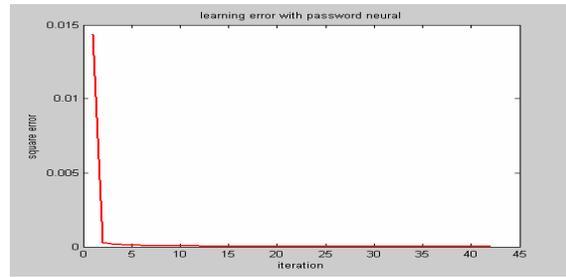

**Figure 5. Learning plot**

Learning has given corresponding to password 'neural'. It's required very few number of iteration to minimize the error below 0.00001,as shown in fig (5). The available trained weights utilize for authentication process. In the test phase four different passwords 'neural', 'mural, 'neurba', 'signal' has taken to verify the process. Its very clear from the result given system performs as it was expected. Zero difference appears for password' neural' while for any other case large difference, which is very good enough for declaring as a unauthentication, as shown in fig (6).

### 7.2 Experiment 2

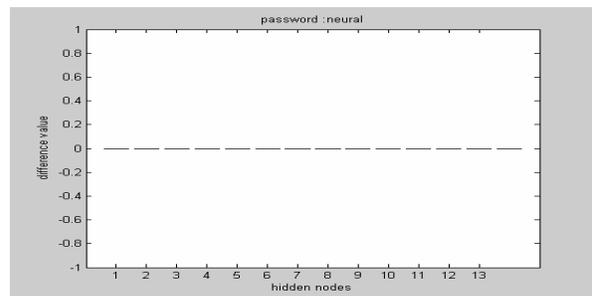

**Figure 6(a) test password: neural**

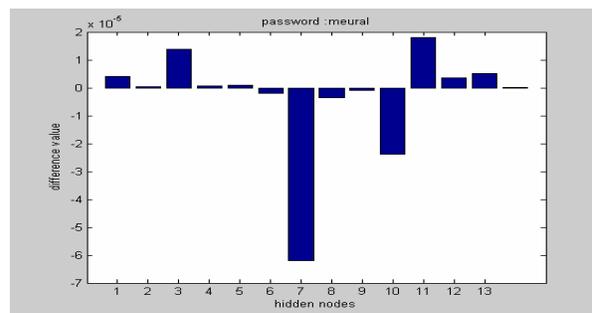

**Figure 6(b) test password: meural**

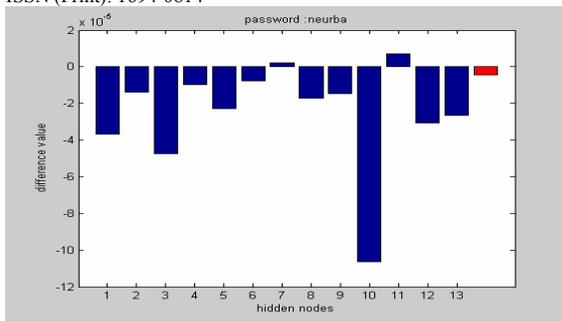

**Figure 6(c) test password: neurba**

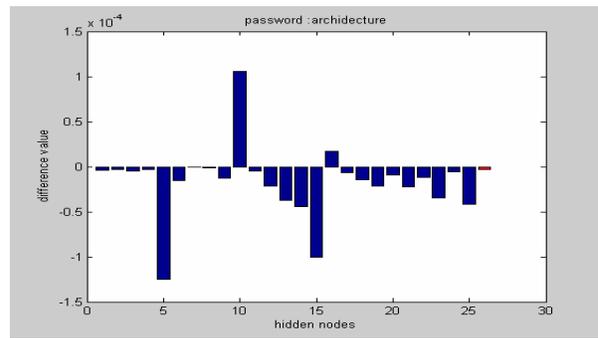

**Figure 8(b) test password: archidecture**

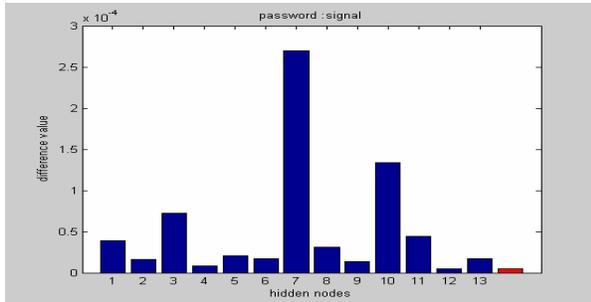

**Figure 6(d) test password: signal**

**(* Last bar with red color represent the final output mapping difference)**

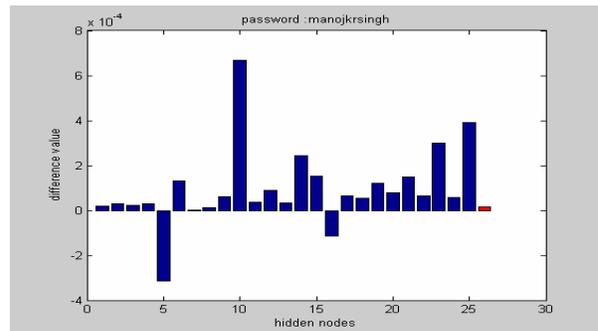

**Figure 8(c) test password: manojkrsingh**

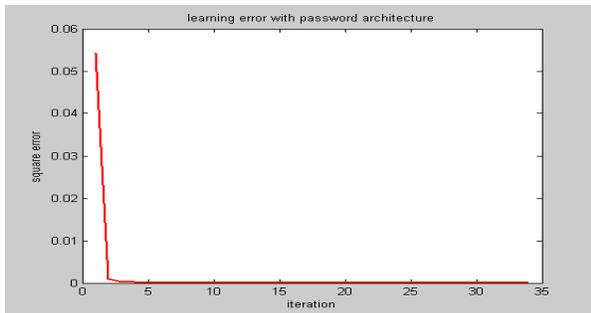

**Figure 7. Learning plot**

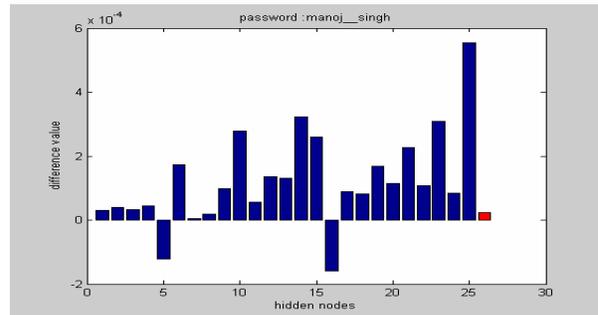

**Figure 8(d) test password: manoj__singh**

## 8 Proposed security system

### 8.1 Hierarchical structure of security

Based upon above discussion a multiplayer, multiparametric security system has developed. In most of the public oriented services there are service provider along with service users. Each party wanted to allocated a security means to protect their resource. Hence for each case a separate password provision can be allocated (if there is no service provider single password can also be applied). To provide the intruder detection several parameters can taken as measuring consideration likes, (1) number of trails (2) length of password (3) time taken to insert the password. These parameters can surely help to define the activity as normal or intrusion. These parameters are forming a hierarchical structure to enhance the speed of computation and saving the

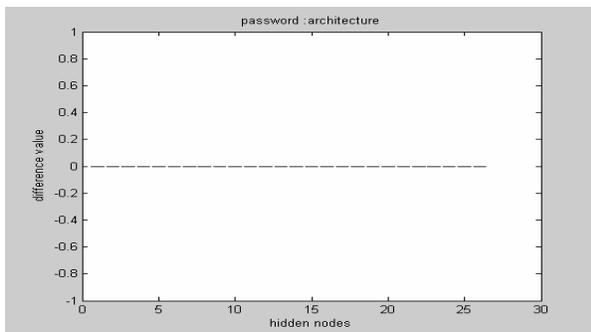

**Figure 8(a) test password: architecture**

unnecessary involvement of other inner layer modules associated with this security system. These three layers logically define the types of activity without knowing the contents of password. Neural network based authentication is the innermost layer, which is taking care of contents available with the password as shown in Fig (9).

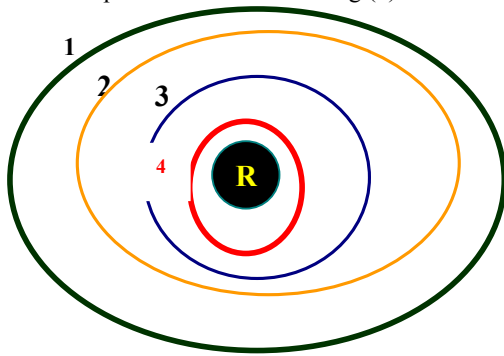

**Figure 9. 1:Trail layer, 2:length layer, 3:Time layer, and 4:ANN layer, R: Resource.**

Any request having permission by all layers only can access the resource otherwise any failure at any layer will be counted as unauthentic activity. Any unauthentic activity will cause to save all the information about the events like time of event, inserted password, and time taken to insert the password. Repeated unauthentic activities (depend upon maximum trail permission allocated for that particular application) result in declaration of intruder in front of system (this situation may also appear if right user have forgotten the password). When intrusion declared, to protect the resource, security environment not allow entering the password further. The other benefit of this facility is, when the right person will try to access the resource, system will not permit to enter the password hence an auto- information mechanism about intrusion available to right user**.**

### 8.2 Block diagram of appearance of Security System from User Side.

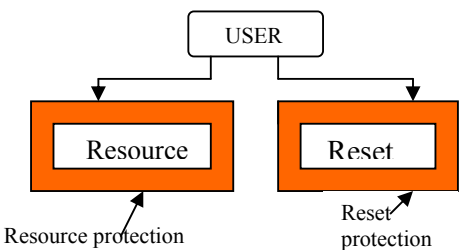

**Figure 10. Front view for user**

for any user there are two path of progress, (1) accessing of resource (2) reset the old password or intrusion file. If user going for resource access, above defined protocol will follow. Reset mode basically contains two facilities namely intrusion activities file or reset the old passwords either by new password or retaining the existing passwords. Picking up of reset mode is a necessary step for right user if there is seize in resource mode. Because reset mode is open for all, hence a high level of security provision has given for this by allocating a reset password along with the two same passwords protecting the resource. In summary to reset the passwords three currently existing passwords must be correctly given by the user. The flow of reset mode is given in fig. (11).

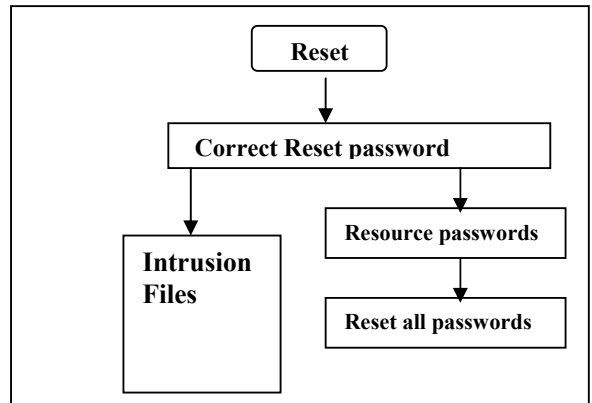

**Figure 11. Reset protocol**

Resetting the passwords with same existing ones is having its own advantage in this system. Even for same password the encrypted information will very different compare to previous existing information. Hence instead of changing the password frequently this is much easy means to change the contents of encrypted information.

### 8.3 Two factor security with physical device

With the association of a user-friendly memory device the level of security with given method can be increase very much. The core reason behind this is availability of two different sets of trained weights, hidden layer weights and output layer weights, from neural network. If hidden layer weights transfer to user memory device and output layer weights stored in server or with resource, resistance with brute force trail of attacks can be increase very much because virtually there are infinite possibility exist in the real value domain. Hence even with simple and small password, high level of security can be achieved. Because with each new training of neural network there are different set of weights appear, there is no harm of using the same password with multi-applications. This will remove the burden of memory aids. This proposed method give the level of security comparatively better and cost effective to even smart card.

### 8.4 Advantages of purposed system

In summary the security system given in this paper having advantages like: -
■ a similar, simple method to design the one-way hash function.
■with new training of same password encrypted information will be different,. Hence instead to changing the password periodically, with same password the security level can be increased very much.
■ free from service provider faith ness circumference.

IJCSI

■ Intrusion detection facility.
■ Hierarchical protection gives optimum use of security model with high processing speed.
■ Multi-user facility from same security environment.
■ Encrypted password information always appear in a set of real value.
■ can be integrating with any kind of resource environment like soft resource or physical resource.
■ with association of external memory device, level of security can achieve a great height and reduce the burden of memory aids.

This project has implemented on MATLAB platform and for experiment it has integrated successfully with various products developed by Manuro tech. research lab.

**Conclusion**
**A new method to design the security system using artificial neural network having the intrusion detection capability too, has given. This design is very efficient & robust, at the same time simple. This design having the domain of application numerous irrespective of their nature. With this design, the password identity based security system will get a new direction of development. There is a very good scope to enhance the level of security by splitting the encrypted information in two parts. One part can be stored with security environment whereas other part can embed in to any user-friendly device. As an extension of this design, keystrokes dynamics inclusion can boost the further strength of the solution.**

**Acknowledgments**

Author is thanking to Reeta Singh for giving tremendous support, and to lovely Manoree who always inspire and sometimes surprised me to see the real meaning of evolution. Last but not the least I am thankful to all of my associates at Manuro tech. research lab.

**IJCSI**

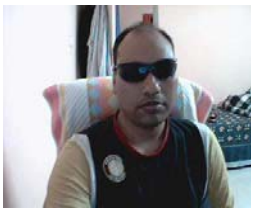

**Manoj kr. Singh,** Currently holding the post of director in manuro tech. research lab. He is actively associated with industry as technology consultant & guiding several research scholars. He is having background of R&D in Nanotechnology, VLSI CAD, Evolutionary computation, Neural network, Advanced signal processing etc.